\documentclass[aps,twocolumn,showpacs]{revtex4}

\usepackage{graphicx}
\include{amssym}
\def\PSfig#1#2{\centerline{\scalebox{#1}{\includegraphics{#2}}}}

\newcommand{\eff}{\mathrm{ef\mbox{}f}}
\newcommand{\ud}{\mathrm{d}}
\newcommand{\dyn}{\mathrm{dyn}}
\newcommand{\crit}{\mathrm{cr}}

\begin{document}

\title{Dynamical chiral symmetry breaking by a magnetic f\mbox{}ield
       \\ and multi-quark interactions}

\author{A. A. Osipov}
\affiliation{Dzhelepov Laboratory of Nuclear Problems, JINR,
         141980 Dubna, Russia}
\author{B. Hiller, A. H. Blin, J. da Provid\^encia}
\affiliation{Centro de F\'{\i}sica Te\'{o}rica, Departamento de F\'{\i}sica
        da Universidade de Coimbra, 3004-516 Coimbra, Portugal}

\begin{abstract}
Catalysis of dynamical symmetry breaking by a constant magnetic
f\mbox{}ield in $(3+1)$ dimensions is considered. We use the three
f\mbox{}lavour Nambu -- Jona-Lasinio type model with 't Hooft and
eight-quark interaction terms. It is shown that the multi-quark
interactions introduce new additional features to this phenomenon:
(a) the local minimum of the ef\mbox{}fective potential catalyzed by
the constant magnetic f\mbox{}ield is smoothed out with increasing
strength of the f\mbox{}ield at the characteristic scale $H\sim
10^{19}\ \mbox{G}$, (b) the multi-quark forces generate independently
another local minimum associated with a larger dynamical fermion mass.
This state may exist even for multi-quark interactions with a
subcritical set of couplings, and is globally stable with respect to a
further increase of the magnetic f\mbox{}ield.
\end{abstract}


\maketitle


It has been shown in a series of papers 
\cite{Lemmer:1989,Klimenko:1991,Krive:1991} that in $(2+1)$ and
$(3+1)$ dimensions a constant magnetic f\mbox{}ield $H\neq 0$
catalyzes the dynamical symmetry breaking leading to a fermion mass 
even at the weakest attractive four-fermion interaction between 
particles, and the symmetry is not restored at any arbitrarily large 
$H$. Soon thereafter it became also clear
\cite{Gusynin:1994,Miransky:1995,Miransky:1996} that the zero-energy
surface of the lowest Landau level (LLL) plays a crucial role in the
dynamics of such fermion pairing. It has been found that the dynamics
of the fermion pairing in the homogeneous magnetic f\mbox{}ield is
essentially $(1+1)$-dimensional, and a deep analogy of this phenomenon
with the dynamics of electron pairing in BCS \cite{Bardeen:1957} has
been stressed. The generated fermion mass, $M_{\dyn}$, turned out to be
much smaller than the Landau gap $\sim\sqrt{|eH|}$.

The existence of a zero-energy surface in the spectrum of a Dirac
particle is ensured for any homogeneous magnetic f\mbox{}ield with a
f\mbox{}ixed direction by a quantum mechanical supersymmetry of the
corresponding second-order Dirac Hamiltonian \cite{Jackiw:1984}.
This aspect of the phenomenon appears to be a quite exceptional
situation and indicates that the dynamical generation of mass is not
so universal as one would expect by extrapolating the results obtained
for homogeneous or unidirectional \cite{Ragazzon:1994} magnetic
f\mbox{}ield prof\mbox{}iles. For instance, it has been demonstrated
by Ragazzon \cite{Ragazzon:1999} that the Nambu -- Jona-Lasinio (NJL)
model \cite{Nambu:1961} minimally coupled to a background magnetic
f\mbox{}ield with variable direction does not possess a massive
phase until the coupling constant exceeds some critical value.
Obviously, in this case one faces the conventional scenario of
dynamical chiral symmetry breaking, where the magnetic f\mbox{}ield 
does not play an essential role.

Conversely, having in mind that homogeneous magnetic f\mbox{}ields can
act as strong catalysts of chiral symmetry breaking, one might ask what
is the ef\mbox{}fect caused by the strong interaction, when higher order
multi-fermion interactions are present. These extensions of the NJL
model are well-known \cite{Eguchi:1976,Volkov:1982,Ebert:1986}, for
instance, the four-quark $U(3)_L\times U(3)_R$ chiral symmetric
Lagrangian together with the $U(1)_A$ breaking 't Hooft six-quark
interactions has been extensively studied at the mean-f\mbox{}ield level
\cite{Bernard:1988,Reinhardt:1988,Weise:1990,Hatsuda:1994}.
Recently it has been also shown \cite{Osipov:2006, Osipov:2007} that
the eight-quark interactions are of vital importance to stabilize the
multi-quark vacuum.

The additional multi-quark forces can af\mbox{}fect the result which
is obtained when only four-fermion interactions are considered. 
We argue, in particular, that the 't Hooft and eight-quark
interactions can modify the theory in such a way that the local 
minimum, catalyzed by the constant magnetic f\mbox{}ield, is smoothed 
out by increasing the strength of the f\mbox{}ield. This is an 
alternative regime to the known one in which the strong magnetic 
f\mbox{}ield cannot wash out the condensate from the LLL. For the 
f\mbox{}irst scenario to become possible it is suf\mbox{}f\mbox{}icient 
that the couplings of multi-quark interactions are chosen such that
the system displays more than one solution of the gap equation at
$H=0$. Actually, the above condition is not a requirement. Even if the 
gap equation has only one nontrivial solution at small $H$, an
increase in the magnetic f\mbox{}ield can induce the formation of a 
second minimum.

The multi-quark dynamics of the extended NJL model is described by
the Lagrangian density
\begin{equation}
\label{efflag}
  {\cal L}_{\eff} =\bar{q}(i\gamma^\mu D_\mu - \hat{m})q
          +{\cal L}_{4q} + {\cal L}_{6q} + {\cal L}_{8q},
\end{equation}
where the gauge covariant derivative $D_\mu$ is equal to $D_\mu =
\partial_\mu +iQA_\mu$ with the external electromagnetic f\mbox{}ield
$A_\mu$ and quark charges $Q=e\cdot\mbox{diag}(2/3,-1/3,-1/3)$. It is
assumed that quark f\mbox{}ields have colour $(N_c=3)$ and f\mbox{}lavour
$(N_f=3)$ indices. The current quark mass, $\hat{m}$, is a diagonal
matrix with elements $\mbox{diag} (\hat{m}_u, \hat{m}_d, \hat{m}_s)$,
which explicitly breaks the global chiral $SU_L(3)\times SU_R(3)$
symmetry of the Lagrangian. We shall neglect this ef\mbox{}fect in the
following assuming that $\hat{m}=0$.

The multi-quark interactions (in the scalar and pseudoscalar channels)
are
\begin{eqnarray}
\label{L4q}
  {\cal L}_{4q} &\!\! =\! & \frac{G}{2}\left[(\bar{q}
  \lambda_aq)^2+ (\bar{q}i\gamma_5\lambda_aq)^2\right], \\
\label{Ldet}
  {\cal L}_{6q} &\!\! =\! &\kappa (\mbox{det}\ \bar{q}P_Lq
                      + \mbox{det}\ \bar{q}P_Rq), \\
  {\cal L}_{8q}&\!\! =\! & {\cal L}_{8q}^{(1)} +
                           {\cal L}_{8q}^{(2)}.
\end{eqnarray}
The $U(3)$ f\mbox{}lavour matrices $\lambda_a,\ a=0,1,\ldots ,8,$
are normalized such that $\mbox{tr} (\lambda_a \lambda_b) =
2\delta_{ab}$. The matrices $P_{L,R}=(1\mp\gamma_5)/2$ are
chiral projectors and the determinant is over f\mbox{}lavour indices,
which are suppressed here. The determinantal interaction breaks
explicitly the axial $U(1)_A$ symmetry \cite{Hooft:1978} and Zweig's
rule. The eight-quark spin zero interactions are given by
\begin{eqnarray}
\label{L1}
   {\cal L}_{8q}^{(1)}&\!\! =\! &
   8g_1\left[ (\bar q_iP_Rq_m)(\bar q_mP_Lq_i) \right]^2, \\
\label{L2}
   {\cal L}_{8q}^{(2)}&\!\! =\! &
   16 g_2 (\bar q_iP_Rq_m)(\bar q_mP_Lq_j)
   (\bar q_jP_Rq_k)(\bar q_kP_Lq_i).
\end{eqnarray}
$G,\, \kappa ,\, g_1,\, g_2$ are dimensionful coupling constants:
$[G]=M^{-2},\ [\kappa ]=M^{-5},\ [g_1]=[g_2]=M^{-8}$ in units
$\hbar =c=1$.

We proceed by calculating the ef\mbox{}fective potential of the
theory, $V(m_u, m_d, m_s)$, in a constant magnetic f\mbox{}ield:
$A_x=-Hy,\, A_y=A_z=0$ (Landau gauge). The arguments, $m_i$, are
simply real parameters; they are not to be identif\mbox{}ied with the
masses of any presumed one-particle states. Instead, we shall use the
capital letter $M_i$ for the point where $V$ takes its local minimum,
which specif\mbox{}ies the masses of constituent quark f\mbox{}ields.

The potential is built of the following two terms
\begin{equation}
\label{efpot}
   V(m_u,m_d,m_s)=V_{st}+V_{S}.
\end{equation}
The f\mbox{}irst contribution results from the many-fermion vertices
of Lagrangian ${\cal L}_{\eff}$, after reducing them to a bilinear form
with help of bosonic auxiliary f\mbox{}ields, and subsequent
integration over these f\mbox{}ields, using the stationary phase
approximation (SPA) method. The specif\mbox{}ic details of these
calculations and the result are given in a recent work
\cite{Osipov:2006}. We obtain
\begin{equation}
\label{effpot1}
     V_{st} = \frac{1}{16}
     \left( 4Gh^2  + \kappa h_uh_dh_s + \frac{3g_1}{2}
     \left(h^2\right)^2 +3g_2 h^4\right),
\end{equation}
where $h^2=\sum_{i=u,d,s}h_i^2$, and $h^4=\sum_{i=u,d,s}h_i^4$. The
functions $h_i$ depend on the coupling constants $G,\kappa, g_1, g_2$
and on the independent variables $\Delta_i=m_i-\hat{m}_i$. To
f\mbox{}ind this dependence one should solve the system of cubic
equations
\begin{equation}
\label{saddle-1}
   Gh_i + \Delta_i +\frac{\kappa}{16}\left. h_jh_k\right|_{j\neq
   k\neq i}
   +\frac{g_1}{4}\ h_i h^2
   +\frac{g_2}{2}\ h_i^3=0.
\end{equation}
In some parameter range the system has only one set of real solutions,
and this guarantees the vacuum state of the theory to be stable
\cite{Osipov:2006}.

The second term on the r.h.s. of eq. (\ref{efpot}) derives from the
integration over the quark bilinears in the functional integral of the
theory in presence of a constant magnetic f\mbox{}ield $H$. As has
been calculated by Schwinger a long time ago \cite{Schwinger:1951}
\begin{equation}
\label{VStot}
     V_{S} = \sum_{i=u,d,s} V_S(m_i, |Q_iH|),
\end{equation}
where
\begin{eqnarray}
   && V_S(m, |QH|)  \\
   && =\frac{N_c}{8\pi^2}\int\limits_0^\infty
   \frac{\ud s}{s^2} e^{-sm^2}\rho (s,\Lambda^2)
   |QH|\coth (s|QH|)+\mathrm{const}. \nonumber
\label{SR}
\end{eqnarray}
Here the cutof\mbox{}f $\Lambda$ has been introduced by subtracting
of\mbox{}f suitable counterterms to regularize the integral at the
lower limit, {\it i.e.}, $\rho (s,\Lambda^2)=1-(1+s\Lambda^2)
e^{-s\Lambda^2}$. For the fermion tadpole this works as the
four-momentum covariant cutof\mbox{}f in the euclidean space:
$\vec{p}\,^2+p_4^2<\Lambda^2$. The unessential constant is chosen such 
that $V_S(0,|QH|)=0$. As a result we obtain
\begin{eqnarray}
   && V_S(m, |QH|)  \nonumber \\
   && =\frac{N_c}{8\pi^2}\left\{\Lambda^2|QH|\left[\ln 2\pi
      - 2\ln\Gamma\left(\frac{\Lambda^2+m^2}{2|QH|}\right)\right]
      \right.\nonumber \\
   && +m^2|QH|\ln\left(1+\frac{\Lambda^2}{m^2}\right) +4(QH)^2
      \nonumber\\
   && \times\frac{\ud}{\ud\nu}\left[ \zeta\left(\nu -1,
   \frac{\Lambda^2+m^2}{2|QH|}\right)-
   \left.\zeta\left(\nu -1, \frac{m^2}{2|QH|}\right)\right]
   \right|_{\nu =0}   \nonumber \\
   && \left. +\frac{\Lambda^4}{2}\left(\ln\frac{\Lambda^2}{2|QH|}
   -\frac{3}{2}\right)-\Lambda^2m^2\right\}.
\label{VS}
\end{eqnarray}
The quantity $\zeta (\nu ,x)$ denotes the generalized Riemann zeta
function \cite{Bateman:1953}.

\begin{figure}[t]
\PSfig{0.45}{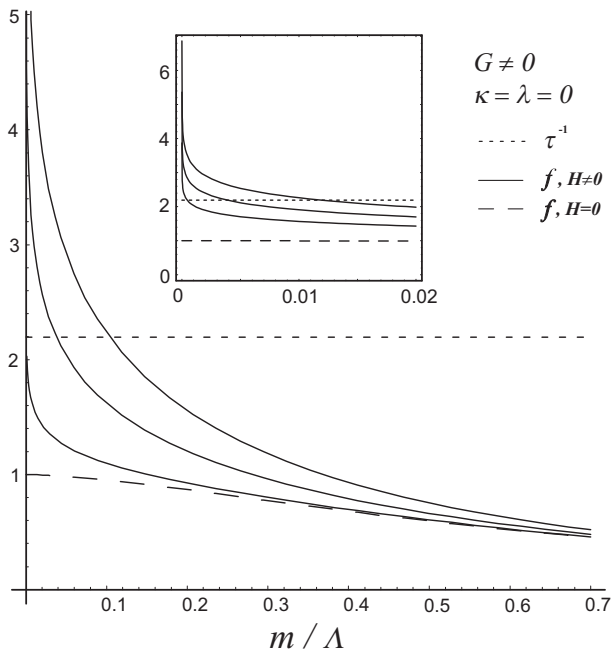}
\caption{\small The l.h.s. (straight short-dashed line) and the r.h.s.
  of eq. (\ref{gap2}) as functions of $m/\Lambda$ for four
  dif\mbox{}ferent values of the magnetic f\mbox{}ield strength $H$:
  full curves (top to bottom) correspond to
  $|QH|\Lambda^{-2}=0.5;\, 0.3;\, 0.1$, and the dashed curve
  to $H=0$. Box insert: close-up of region around origin with solid
  lines for $|QH|\Lambda^{-2}=0.2;\, 0.15;\, 0.1$ (top to bottom).
  }
\label{fig1}
\end{figure}

We shall now illustrate the procedure which will be employed in the
following, by considering f\mbox{}irst the simple $SU(3)$ f\mbox{}lavour
limit for the situation in which $\hat m=0$ and $\kappa =g_1=g_2=0$.
For the purpose of illustration, we ignore  the charge
dif\mbox{}ference of $u$ and $d,s$ quarks in the remaining. The
averaged common charge $|Q|=|4e/9|$ will be used. In this case one 
obtains the potential $V(m)=N_f(m^2/4G+V_S(m, |QH|))$. One sees that 
the gap equation, $\ud V(m)/ \ud m = 0$,  has always a trivial 
solution, $m=0$. The nontrivial solution is contained in the 
equation
\begin{eqnarray}
\label{gap2}
   && \frac{2\pi^2}{G\Lambda^2N_c} = f(m^2;\Lambda,|QH|) \equiv
   \psi\left(\frac{\Lambda^2+m^2}{2|QH|}\right)
   - \frac{|QH|}{\Lambda^2} \nonumber \\
   &&\!\times\!\left[\ln\!\left(\! 1+\frac{\Lambda^2}{m^2}\right)
     \! -\frac{\Lambda^2}{\Lambda^2+m^2}  +2\ln
   \frac{ \Gamma\left( \frac{\Lambda^2+m^2}{2|QH|} \right)
   }{\Gamma \left( \frac{m^2}{2|QH|} \right)}\right]\! ,
\end{eqnarray}
where $\psi (x)=\ud\ln\Gamma (x)/\ud x$ is the Euler dilogarithmic
function. This equation has a solution at all $G>0$, if $H\neq 0$. 
F\mbox{}ig. \ref{fig1} illustrates this important result for 
$G \Lambda^2=3$. One sees that in absence of the magnetic f\mbox{}ield 
the system is in the subcritical regime of dynamical symmetry breaking.
The introduction of a constant f\mbox{}ield, however small it might
be, changes radically the dynamical symmetry breaking pattern, due to
the singular behaviour of the r.h.s. of eq. (\ref{gap2}) close to the
origin: the right and left hand sides will always intersect and the 
value of $m$ where this happens is a minimum of the ef\mbox{}fective 
potential.

Let's look at these dif\mbox{}ferent regimes more closely. As $H\to
0$, we recover the well-known NJL model gap equation
\begin{equation}
\label{stgap}
   1=\frac{GN_c}{2\pi^2}\left(J_0(m^2)
    +\frac{|QH|^2\Lambda^2}{3m^2(\Lambda^2+m^2)}+\ldots\right),
\end{equation}
where $J_0(m^2)=\Lambda^2 -m^2\ln\left(1+\Lambda^2/m^2\right).$
Eq. (\ref{stgap}) at $H=0$ admits a nontrivial solution only if
$\tau >1$, where $\tau =G\Lambda^2N_c/2\pi^2$. This determines the
critical value $G_{\crit}=2\pi^2/\Lambda^2N_c$.

At $m^2/\Lambda^2\ll 1$ the r.h.s. of eq. (\ref{gap2}) is
\begin{equation}
         - \frac{|QH|}{\Lambda^2}\ln\left(
           \frac{m^2}{\Lambda^2}\right)+v(\xi )
           +{\cal O}\left(\frac{m^2}{\Lambda^2}\right).
\label{rhs}
\end{equation}
Here the function $v(\xi )$ of the argument $\xi =\Lambda^2/2|QH|$, is
given by
\begin{equation}
\label{nlo}
   v(\xi )=\frac{1}{2\xi}\left[1-2\ln\Gamma\left(\xi
   +1\right)\right] + \psi\left(\xi\right).
\end{equation}
This is a monotonically increasing function on the interval $0<\xi
<\infty$; $v(\xi )=0$ at the point $\xi\simeq 1.12$; the asymptotic
behaviour is
\begin{eqnarray}
\label{asinf}
   v(\xi )&\!\sim\!&1-\frac{1}{2\xi}\ln (2\pi\xi),\quad
                    (\xi\to\infty ), \\
   v(\xi )&\!\sim\!&-\gamma -\frac{1}{2\xi}, \quad (\xi\to 0),
\label{as0}
\end{eqnarray}
where $\gamma\simeq 0.577$ is the Euler's constant.

In the approximation considered one f\mbox{}inds the solution of eq.
(\ref{gap2})
\begin{equation}
\label{dynm1}
   M_{\dyn}=\Lambda \exp\left[-\xi\left(
            \frac{1}{\tau}-v(\xi )\right)\right].
\end{equation}

To discuss the physical content of this result, we recall that the
energy spectrum of relativistic fermions in a constant magnetic
f\mbox{}ield $H$ contains Landau levels
\begin{equation}
   E_n(p_z)=\pm\sqrt{\hat m^2+2|QH|n+p_z^2}, \quad n=0,1,2,\ldots
\end{equation}
with $p_z$ denoting the projection of the 3-momentum on the $z$-axis,
{\it i.e.,} along the magnetic f\mbox{}ield. If the fermion mass $\hat
m$ goes to zero, as in the present case, there is no energy gap
between the vacuum and the LLL. Thus the integer part of $\xi +1$
gives approximately the number of Landau levels taken into account.

The f\mbox{}irst term in eq. (\ref{rhs}) has a clearly def\mbox{}ined
two-dimensional origin, given the logarithmic dependence on the
cutof\mbox{}f in the corresponding gap equation
\begin{equation}
\label{gap3}
   1=-\frac{GN_c}{2\pi^2}|QH|\ln\left(\frac{m^2}{\Lambda^2}\right)
\end{equation}
and, therefore, in the condensate (compare with eq. (\ref{stgap})).
Such behaviour is associated with the $(1+1)$-dimensional dynamics of
the fermion pairing on the energy surface $E_0=0$ of the LLL
\cite{Miransky:1995}. As long as this term dominates over the second
term, $v(\xi )$ in (\ref{rhs}), one concludes that the condensate is
mainly located on the LLL. Actually this condition is fulf\mbox{}illed
nearly everywhere at $\tau < 1$. Indeed, this is obvious for $\xi =1$,
since $v(1) = 1/2+\psi (1) = 1/2 -\gamma\simeq -0.08$ is small compared
with $1/\tau$. For $\xi <1$ we come to the same conclusion after
considering the asymptotics of the second term (\ref{as0}). The other
formula, (\ref{asinf}), can be used to show that the above statement
is also true for $\xi >1$, except near the critical region $\tau\to
1-0$, where $v(\xi )$ dominates; then the condensate spreads over many 
Landau levels.

In this special case it is possible to f\mbox{}ind an analytical
solution. Indeed, using (\ref{asinf}) in eq. (\ref{rhs}) we obtain
\begin{equation}
\label{rhs3}
   1-\frac{|QH|}{\Lambda^2}\ln\left(\frac{\pi m^2}{|QH|}\right)
   + {\cal O}\left(\frac{m^2}{\Lambda^2}, \frac{4|QH|^2}{\Lambda^4}
   \right).
\end{equation}
To progress  further we suppose that the two following small
variables are of the same order
\begin{equation}
\label{cond}
  \frac{m^2}{\Lambda^2}\sim
  \left(\frac{|QH|}{\Lambda^2}\right)^2\sim\epsilon .
\end{equation}
Then it follows immediately that the term with the logarithm is of
order $\sqrt{\epsilon}\ln\sqrt{\epsilon}$ and goes to zero, when
$\epsilon\to 0$. Thus, the gap equation
\begin{equation}
\label{tau}
   1-\frac{1}{\tau}= \frac{|QH|}{\Lambda^2}
   \ln\left(\frac{\pi m^2}{|QH|}\right) + {\cal O}(\epsilon )
\end{equation}
is valid only in the region near the critical value $\tau\to 1-0$.
The closer $\tau$ to $1$, the smaller is $\epsilon$; Landau levels
approach a continuun distribution, and a condensate occupies many
levels. The physical reason for the changes found in the behaviour
of the condensate is the strength of the four-fermion interaction
which becomes essentially important here. The corresponding solution
is
\begin{equation}
   M_{\dyn}^2=\frac{|QH|}{\pi}\exp\left[-\frac{\Lambda^2}{|QH|}
             \left(\frac{1}{\tau}-1\right)\right].
\end{equation}

Note that the near-critical regime found here, dif\mbox{}fers from the
result of Ref. \cite{Miransky:1995}, being driven by a quadratic
dependence on the cutof\mbox{}f, eq. (\ref{tau}); this is tantamount of
having a $(3+1)$-dimensional dynamics of fermion pairing.

Let us return now again to the three f\mbox{}lavour case with $\kappa,
g_1, g_2\neq 0$. In the simplest case with the octet f\mbox{}lavour
symmetry, where current quarks have equal masses
$\hat{m}_u=\hat{m}_d=\hat{m}_s$, which we set again zero, the system
(\ref{saddle-1}) reduces to a cubic equation with respect to
$h\equiv h_u=h_d=h_s$
\begin{equation}
\label{cubeq1}
   h^3 + \frac{\kappa}{12\lambda}\, h^2
         +\frac{4G}{3\lambda}\, h + \frac{4m}{3\lambda}=0
\end{equation}
with $\lambda =g_1+(2/3)g_2$. This cubic equation has one real root, if
(see \cite{Osipov:2007} for more details)
\begin{equation}
\label{stabcond}
    \frac{G}{\lambda}>\left(\frac{\kappa}{24\lambda}\right)^2.
\end{equation}
Assuming that the couplings fulf\mbox{}ill condition (\ref{stabcond}),
we f\mbox{}ind a single valued function $h(m)$ from eq. (\ref{cubeq1}).

\begin{figure}[t]
\PSfig{0.5}{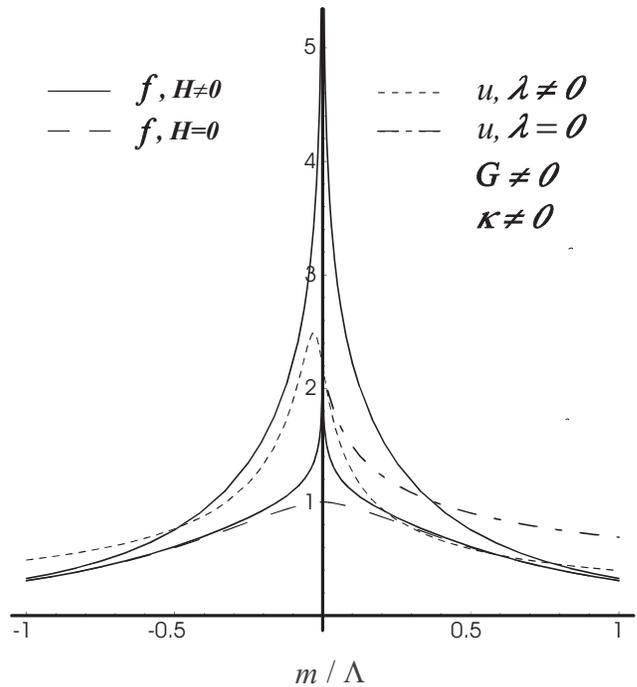}
\caption{\small The l.h.s. and the r.h.s. of eq. (\ref{gap5}) as
   functions of $m/\Lambda$ with $f(m^2;\Lambda,|QH|)$ shown for
   $|QH|\Lambda^{-2}=0;\, 0.1;\, 0.5$. The l.h.s. with $\lambda\ne 0$
   (short-dashed line) is shifted with respect to the ordinate axis
   due to the 't Hooft interaction, the range of values for the
   $\lambda =0$ curve (dash-dotted) is limited (see discussion in text).
   The value of $G\Lambda^2=3$ is the same as in f\mbox{}ig. \ref{fig1}. 
   Here the six- and eight-quark couplings are taken to be
   $\kappa\Lambda^5=-10^3$ and $\lambda \Lambda^8= 3.67\cdot 10^3$
   (or $\lambda=0$), respectively.  }
\label{fig2}
\end{figure}

Considering that most of the investigations have been using
multi-quark Lagrangians without the stabilizing eight-quark
interactions we make a short digression to discuss the case with
$\lambda=0$ (see details in {\it e.g.} \cite{Osipov:2006,Osipov1:2006}).
In this case eq. (\ref{cubeq1}) is quadratic with a regular and a
singular solution as $\kappa\rightarrow 0$, {\it i.e.,}
\begin{equation}
   h^{(1,2)} = -\frac{8 G}{\kappa} \left( 1\mp\sqrt{1-
   \frac{\kappa m}{4 G^2}} \right).
\end{equation}
In SPA this leads to an unstable ef\mbox{}fective potential. In the
commonly used mean f\mbox{}ield approximation, which discards the
singular solution, the ef\mbox{}fective potential is metastable and
the region $4 G^2< \kappa m$ leads to complex values for $h$. This 
translates to a restriction for the admissible values that
the l.h.s. of eq. (\ref{gap5}) can assume, shown as dash-dotted line
in f\mbox{}ig. \ref{fig2}, discussed below. Furthermore we f\mbox{}ind 
that qualitatively the symmetry breaking pattern is the same as in 
presence of the eight-quark interactions, but the occurrence of two 
minima requires higher values of $|\kappa|$, as compared to the case 
with $\lambda\ne 0$ ($\sim 2.2 \kappa$ for the parameter set of 
f\mbox{}ig. \ref{fig2}). From now on we shall consider only the case
which fulf\mbox{}ills the stability requirement, provided by eq. 
(\ref{stabcond}).

The nontrivial solutions for the mass of the fermion f\mbox{}ield
in a constant magnetic f\mbox{}ield are determined by the equation
\begin{equation}
\label{gap5}
  -\frac{2\pi^2h(m)}{\Lambda^2N_cm}= f(m^2;\Lambda,|QH|).
\end{equation}
Comparing this result with eq. (\ref{gap2}), one sees that only the
l.h.s. is changed. The six- and eight-quark interactions have
modif\mbox{}ied it in such a way that now we get a function $h(m)/m$
instead of the former constant term involving only the coupling of
four-quark interactions, $-1/G$. The l.h.s. of eq. (\ref{gap5}),
abbreviated by $u$ in f\mbox{}ig. 2, has now a bell-shaped form
(short-dashed line), to be compared with the horizontal line of
f\mbox{}ig. 1. Note, that $h(m)/m =-1/G+{\cal O}(m)$, {\it i.e.}, the
bell-shaped curve crosses the ordinate axis at the same point as the
former straight line (for the same value of $G\Lambda^2$). The r.h.s. 
of the eq. (\ref{gap5}) is again represented by the long-dashed curve 
($H=0$ case) and by the full lines for the f\mbox{}inite $H$ cases: 
$|QH|\Lambda^{-2}=0.1,\, 0.5$. As mentioned before, these are not 
altered by the couplings $G,\,\kappa,\,\lambda$. The intersection
points of the l.h.s. with the r.h.s. curves yield the nontrivial 
solutions of the gap equation: one sees that either one or three 
solutions can be found for $m>0$. If eq. (\ref{gap5}) has no solutions 
at $H=0$ we say that the set of couplings $G,\,\kappa,\, \lambda$ are 
subcritical. It is said to be overcritical in the opposite case. Note 
that the overcritical set may contain $G<G_\crit$.

The trivial solution, $m=0$, corresponds to the point where the
potential, $V(m)$, reaches its local maximum. Indeed, the second
derivative
\begin{equation}
\label{sder}
   \lim_{m\to 0} \frac{d^2V(m)}{dm^2}=\lim_{m\to 0}
   \frac{N_c|QH|}{2\pi^2}\ln\frac{|m|\Lambda}{2|QH|}=-\infty
\end{equation}
is negative here. This is the general mathematical reason for the
phenomenon known as magnetic catalysis of dynamical f\mbox{}lavour
symmetry breaking. The logarithmically divergent negative result
ensures that this phase transition always takes place, if $H\neq 0$.
This does not depend on the details related with the multi-quark
dynamics, {\it i.e.}, the result is true even
for free fermions in a constant magnetic f\mbox{}ield.

What is really sensitive to the multi-quark dynamics is the local
minima structure of the theory. Let us recall that in the theory
with just four-fermion interactions the ef\mbox{}fective potential has
only one minimum at $m>0$, and this property does not depend on the
strength of the f\mbox{}ield $H$. We have demonstrated this in
f\mbox{}ig. \ref{fig1}. In the theory with four-, six-, and eight-quark
interactions one can f\mbox{}ind either one or two local minima at
$m>0$. The result depends on the strength of the magnetic f\mbox{}ield
$H$, and couplings $G,\,\kappa,\,\lambda$. We illustrate these two
cases in f\mbox{}ig. \ref{fig2}. Namely, the upper full curve (r.h.s. 
of eq. (\ref{gap5}) for $|QH|\Lambda^{-2}=0.5$) has only one intersection
point with the bell-shaped curve $u$ (l.h.s. of eq. (\ref{gap5}) for
$G\Lambda^2=3,\,\kappa\Lambda^5=-1000,\,\lambda\Lambda^8=3670$). This
point corresponds to a single vacuum state of the theory. The other
full curve (r.h.s. of eq. (\ref{gap5}) for $|QH|\Lambda^{-2}=0.1$)
has three intersections with the same curve $u$. These intersections,
successively, correspond to a local minimum, a local maximum and
a further local minimum of the potential.

\begin{figure}[t]
\PSfig{0.45}{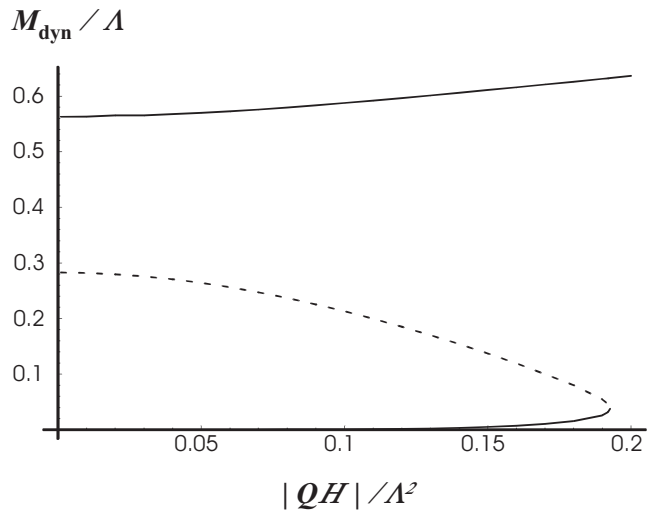}
\caption{\small The dimensionless dynamical mass $M_\dyn/\Lambda$
   as a function of the dimensionless magnetic f\mbox{}ield
   $|QH|/\Lambda^2$. The full lines are minima, the dashed line 
   maxima. Up to $|QH|/\Lambda^2=0.084$ the smaller $M_\dyn/\Lambda$ 
   correponds to the deeper minimum of the potential, from this value 
   on the larger solution becomes the stable configuration.}
\label{fig3}
\end{figure}

It is interesting to note that the f\mbox{}irst minimum catalyzed by
a constant magnetic f\mbox{}ield (that is, a slowly varying
f\mbox{}ield) is then smoothed out with increasing $H$. It ceases to
exist at some critical value of $|QH|\Lambda^{-2}$, from which on only
the large $M_{\dyn}$ solution survives. This is shown in f\mbox{}ig. 
\ref{fig3}, for the parameter set of f\mbox{}ig. \ref{fig2}. This 
process is accompanied by a sharp increase in depth of the 
ef\mbox{}fective potential at the second minimum, especially if we had 
at the beginning the opposite ordering, {\it i.e.}, $V(M_1)<V(M_2)$. 
The reasons for such a synchronized behaviour are the following two 
facts. The f\mbox{}irst one is eq. (\ref{sder}), which teaches us that 
the only way to wash out the f\mbox{}irst minimum is by lowering the 
barrier between this state and the second minimum. The other fact is 
the observation that the second minimum is unremovable, because the 
asymptotic behaviour of the functions in eq. (\ref{gap5}) is such that 
the l.h.s. dominates over the r.h.s. at large $m/\Lambda$. This can
also be understood from f\mbox{}ig. \ref{fig2}, where one sees that 
the r.h.s. of the gap equation with $H\ne 0$ approaches the $H=0$
curve from above.

To discuss the physical content of the phenomenon just described one
should f\mbox{}ix the characteristic scale $\Lambda$. We assume that
this value is determined by the problem under consideration. Its
choice can also be motivated by the number of Landau levels to be
considered.

In the region $m^2/\Lambda^2\ll 1$ the four-fermion interaction
dominates the behaviour of the system. Since their coupling strength
is small, $G<G_\crit$, the massless fermions behave like almost free
particles moving in a weak external magnetic f\mbox{}ield, with access
to a large number of Landau levels, $\xi\gg 1$. This f\mbox{}ield
catalyzes the process of fermion-antifermion pairing on the energy
surface $E_0=0$ of the LLL. The f\mbox{}irst minimum localized at
$m^2/\Lambda^2\ll 1$ is exactly formed by such a $(1+1)$-dimensional
condensate. If six- and eight-fermion forces would not act on the
system, this ground state would be stable: our formulae as well as the
result of paper \cite{Miransky:1995} show clearly that a slow increase
of the strength $H$ does not wash out the condensate from the energy
surface $E_0=0$ of the LLL. However, when the six- and eight-fermion
interactions are present, a slowly increasing magnetic f\mbox{}ield
destroys f\mbox{}inally this ground state. The new condensate has a
$(3+1)$-dimensional structure similar in every respect to the standard
NJL case with broken chiral symmetry at $H=0$, {\it i.e.}, when the
condensate spreads over many single fermion states. This is because
the increasing magnetic f\mbox{}ield enlarges the dynamical fermion
mass, and scales of order $m/\Lambda\sim 1$ become relevant. At these
scales the 't Hooft and eight-quark interactions push the system to a
new regime, where the fermions are not anymore free-like particles: 
they interact strongly with each other and this interaction changes 
the fermionic spectrum and the structure of the ground state in an 
essential way, with all the above mentioned consequences.

Thus we have obtained not only a correct description of the
well-known physics related with the LLL, but have found also a clear
signature for the possibly important role played by 't Hooft and
eight-quark interactions. Namely, in the presence of these
interactions the magnetic f\mbox{}ield can change the condensation
zone from the zero-energy surface of the LLL to a wide region, spread
over many Landau levels and vice versa. One can expect that hard gamma
emissions accompany this process.

What is the characteristic scale of the magnetic f\mbox{}ields which
can induce such a transition? We obtained the value $H=7.3\cdot 10^{13}\,
\Lambda^2\,\mbox{Gauss/MeV$^2$}$ which actually depends on the
cutof\mbox{}f involved in the problem. For instance, in hadronic
matter it is probably safe to assume that $\Lambda\simeq 800\,
\mbox{MeV}$, leading to $H=4.7\cdot 10^{19}\,\mbox{G}$.

One can indicate several potentially interesting areas where this
ef\mbox{}fect may f\mbox{}ind applications. One of them is connected
with the recent studies of compact stellar objects in presence of strong
magnetic f\mbox{}ields, in particular the young neutron stars,
magnetars \cite{Duncan:1992}. The surface magnetic f\mbox{}ields are
observed to be $\geq 10^{15}\,\mbox{G}$, but actually they can be even
much higher in the core region. The other area is connected with the
electroweak phase transition in the early Universe \cite{Olesen:1992},
where the strength of magnetic f\mbox{}ields can reach
$H\sim 10^{24}\,\mbox{G}$.

\vspace{0.5cm}
{\bf Acknowledgements}
This work has been supported in part by grants provided by
Funda\c c\~ao para a Ci\^encia e a Tecnologia, POCI/FP/63412/2005,
POCI/FP/63930/2005. This research is part of the EU integrated
infrastructure initiative Hadron Physics project under contract
No.RII3-CT-2004-506078.



\begin{thebibliography}{99}

\bibitem{Lemmer:1989} S. P. Klevansky and R. H. Lemmer, Phys. Rev.
      {\bf 39}, 3478 (1989).
\bibitem{Klimenko:1991} K. G. Klimenko, Theor. Math. Phys. {\bf 89},
      211 (1991); K. G. Klimenko, Theor. Math. Phys. {\bf 90}, 3
      (1992); A. S. Vshivtsev, K. G. Klimenko and B. V. Magnitsky,
      Theor. Math. Phys. {\bf 106}, 390 (1996).
\bibitem{Krive:1991} I. V. Krive and S. A. Naftulin, Sov. J. Nucl.
      Phys. {\bf 54}, 897 (1991); I. V. Krive and S. A. Naftulin,
      Phys. Rev. D {\bf 46}, 2737 (1992).
\bibitem{Gusynin:1994} V. P. Gusynin, V. A. Miransky, I. A. Shovkovy,
      Phys. Rev. Lett. {\bf 73}, 3499 (1994).
\bibitem{Miransky:1995} V. P. Gusynin, V. A. Miransky, I. A. Shovkovy,
      Phys. Lett. B {\bf 349}, 477 (1995); Phys. Rev. D {\bf 52}, 4718
      (1995).
\bibitem{Miransky:1996} V. P. Gusynin, V. A. Miransky, I. A. Shovkovy,
      Nucl. Phys. B {\bf 462}, 249 (1996).
\bibitem{Bardeen:1957} J. Bardeen, L. N. Cooper and J. R. Schrief\mbox{}fer,
      Phys. Rev. {\bf 108}, 1175 (1957).
\bibitem{Jackiw:1984} R. Jackiw, Phys. Rev. D {\bf 29}, 2375 (1984);
      A. Barducci, R. Casalbuoni, and L. Lusanna, Nuovo Cimento A {\bf
      35}, 377 (1976).
\bibitem{Ragazzon:1994} R. Ragazzon, Phys. Lett. B {\bf 334}, 427
      (1994); G. Dunne and T. Hall, Phys. Rev. D {\bf 53}, 2220 (1996).
\bibitem{Ragazzon:1999} R. Ragazzon, Phys. Rev. D {\bf 59}, 065006 (1999).
\bibitem{Nambu:1961} Y. Nambu and G. Jona-Lasinio, Phys. Rev. {\bf
      122}, 345 (1961); {\bf 124}, 246  (1961);
      V. G. Vaks and A. I. Larkin, Zh. \'{E}ksp. Teor. Fiz. {\bf 40},
      282 (1961).
\bibitem{Eguchi:1976} T. Eguchi, Phys. Rev. D {\bf 14}, 2755 (1976);
      K. Kikkawa, Progr. Theor. Phys. {\bf 56}, 947 (1976).
\bibitem{Volkov:1982} M. K. Volkov and D. Ebert, Sov. J. Nucl. Phys.
      {\bf 36}, 736 (1982);
      D. Ebert and M. K. Volkov Z. Phys. C {\bf 16}, 205 (1983).
\bibitem{Ebert:1986} M. K. Volkov, Ann. of Phys. {\bf 157}, 282 (1984);
      A. Dhar and S. Wadia, Phys. Rev. Lett. {\bf 52}, 959 (1984);
      A. Dhar, R. Shankar and S. Wadia, Phys. Rev. D {\bf 31}, 3256
      (1985);
      D. Ebert and H. Reinhardt, Nucl. Phys. B {\bf 271}, 188 (1986);
      C. Sch\"uren, E. R. Arriola and K. Goeke, Nucl. Phys. A {\bf 547},
      612 (1992);
      J. Bijnens, C. Bruno and E. de Rafael, Nucl. Phys. B {\bf 390},
      501 (1993), hep-ph/9206236;
      V. Bernard, A. A. Osipov and U.-G Mei\ss ner, Phys. Lett. B {\bf
      324}, 201 (1994), hep-ph/9312203;
      V. Bernard, A. H. Blin, B. Hiller, Yu. P. Ivanov, A. A. Osipov
      and U.-G Mei\ss ner, Ann. of Phys. {\bf 249}, 499 (1996),
      hep-ph/9506309.
\bibitem{Bernard:1988} V. Bernard, R. L. Jaf\mbox{}fe and U.-G.
      Meissner, Phys. Lett. B {\bf 198}, 92 (1987);
      V. Bernard, R. L. Jaf\mbox{}fe and U.-G. Meissner, Nucl. Phys. B
      {\bf 308}, 753 (1988).
\bibitem{Reinhardt:1988} H. Reinhardt and R. Alkofer, Phys. Lett. B
      {\bf 207}, 482 (1988).
\bibitem{Weise:1990} S. Klimt, M. Lutz, U. Vogl and W. Weise,
      Nucl. Phys. A {\bf 516}, 429 (1990);
      U. Vogl, M. Lutz, S. Klimt and W. Weise, Nucl. Phys. A {\bf
      516}, 469 (1990);
      U. Vogl and W. Weise, Progr. Part. Nucl. Phys. {\bf 27}, 195
      (1991).
\bibitem{Hatsuda:1994}
      S. P. Klevansky, Rev. Mod. Phys. {\bf 64}, 649 (1992);
      T. Hatsuda and T. Kunihiro, Phys. Rep. {\bf 247}, 221 (1994),
      hep-ph/9401310.
\bibitem{Osipov:2006} A. A. Osipov, B. Hiller, J. da Provid\^encia,
      Phys. Lett. B {\bf 634}, 48 (2006), hep-ph/0508058.
\bibitem{Osipov:2007} A. A. Osipov, B. Hiller, A. H. Blin, J. da
      Provid\^encia, Ann. of Phys. (N.Y.) (2006),
      doi:10.1016/j.aop.2006.08.004, hep-ph/0607066.
\bibitem{Hooft:1978}
      G. 't Hooft, Phys. Rev. D {\bf 14} (1976) 3432;
      Erratum: {\it ibid} D {\bf 18} (1978) 2199.
\bibitem{Schwinger:1951} J. Schwinger, Phys. Rev. {\bf 82}, 664
      (1951).
\bibitem{Bateman:1953} H. Bateman, A. Erdelyi, Higher Transcendental
      Functions, Mc Graw-Hill Book Company, Inc. 1953.
\bibitem{Osipov1:2006} A. A. Osipov, B. Hiller, V. Bernard,
      A. H. Blin, Ann. of Phys. {\bf 321} (2006) 2504,
      hep-ph/0507226;
      {\it idem} SIGMA {\bf 2} (2006) 026, hep-ph/0602165.
\bibitem{Duncan:1992} R. C. Duncan, C. Thompson, Astrophys. J. Lett.
      {\bf 392}, L9 (1992); C. Thompson, R. C. Duncan, Astrophys. J.
      {\bf 408}, 194 (1993); {\it ibid} {\bf 473}, 322 (1996);
      C. Kouveliotou, {\it et al.,} Nature {\bf 391}, 235 (1999);
      K. Hurley, {\it et al.,} Astrophys. J. {\bf 442}, 111 (1999);
      P. M. Woods, {\it et al.,} Astrophys. J. Lett. {\bf 519}, L139
      (1999).
\bibitem{Olesen:1992} T. Vachaspati, Phys. Lett. B {\bf 265}, 258
      (1991); P. Olesen, Phys. Lett. B {\bf 281}, 300 (1992).
\end{thebibliography}
\end{document}